\documentclass[aps,twocolumn,showpacs,preprintnumbers,prb,amsmath, amssymb,amsfonts,superscriptaddress,floatfix]{revtex4}

\usepackage{amsfonts}
\usepackage{amssymb}
\usepackage{amsmath}
\usepackage{ulem}
\usepackage{color}
\usepackage{latexsym}
\usepackage{graphicx}
\usepackage{subfigure}
\usepackage{graphics}
\usepackage{floatflt}
\usepackage{epsfig}
\usepackage{overpic}
\usepackage{soul} 
\usepackage{varioref,xr-hyper} 
\usepackage[latin1]{inputenc}
\usepackage{mathbbol}
\usepackage{makecell}

\definecolor{blue}{rgb}{0.0,0.35,1.0}

\newcommand{\be}{\begin{equation}}
\newcommand{\ee}{\end{equation}}
\newcommand{\bea}{\begin{eqnarray}}
\newcommand{\eea}{\end{eqnarray}}

\def\lb{\label}

\def\ra{\rightarrow}

\def\Ra{\Rightarrow}

\newcount\bozza \bozza=0
\ifnum\bozza=1
\newdimen\shift \shift=-1.0truecm
\def\lb#1{%
{\label{#1}\rlap{\kern\shift{$\scriptstyle#1$}}}}
\else\def\lb#1{\label{#1}} \fi

\begin{document}
\title{Spin-Based Modeling of Perception as Emergent from contextualized Internal Evaluation}

\author{Laura Fanfarillo}
\affiliation{Istituto dei Sistemi Complessi (ISC-CNR), Via dei Taurini 19, I-00185 Rome, Italy}
\author{Gustavo Diez}
\affiliation{Nirakara Lab, Institute of Research and Cognitive Science, Madrid, Spain}
\author{Victor G\'omez Mayordomo}
\affiliation{Department of Neurology, Institute of Neuroscience, Hospital Universitario Vithas Madrid La Milagrosa, Vithas Hospital Group}
\author{Miguel Bosch}
\affiliation{Instituto de Ciencia de Materiales de Madrid,
ICMM-CSIC, Cantoblanco, E-28049 Madrid, Spain}
\author{J. Ricardo Arias-Gonzalez}
\affiliation{Nanophotonics Technology Center, Universitat Polit\`ecnica de Val\`encia, Camino de Vera s/n, 46022 Valencia, Spain}
\author{Bel\'en Valenzuela}
\affiliation{Instituto de Ciencia de Materiales de Madrid, 
ICMM-CSIC, Cantoblanco, E-28049 Madrid, Spain}

\date{\today}


\begin{abstract}{
We develop a microscopic model of perception of an interoceptive sensation in which spin-like variables encode an organism's internal evaluation of embodied vital norms associated with the sensation. Spins can take positive, negative, or neutral values.
These local evaluations interact on a lattice embedded in the environmental context, and their collective configuration gives rise to a macroscopic perceptual state. 
By applying a coarse-graining procedure to a family of symmetric spin models, we derive a macroscopic Landau-type functional that makes explicit the mechanism by which key phenomenological features of perception emerge from microscopic evaluative interactions. 
A central result is that the inclusion of a neutral evaluative state fundamentally alters the structure of the perceptual landscape, enhancing entropy, lowering the critical threshold, and increasing sensitivity to contextual input. These results establish a principle link between microscopic vital norms evaluations and large-scale perceptual organization of a sensation, offering a flexible framework to relate perceptual regulation and neurobiological modeling. 
The model integrates notions of neuroscience and cognitive science through the formalism of condensed matter field theory, providing a new theoretical language for describing the emergence of perceptual organization from microscopic evaluative interactions.
}\end{abstract}

\maketitle 
\section{Introduction}

Complex organisms exhibit behavior that is coordinated across multiple levels of organization, from microscopic degrees of freedom to macroscopic patterns. Understanding how these scales interact is central to any theory of life and mind~\cite{Longo:2011ab,Levin:2023aa,EmbodiedMindVarela92}. In condensed matter field theory~\cite{Altland:2010aa}, such multiscale interactions are particularly well understood near phase-transition points, where macroscopic modes emerge from the collective behavior of microscopic constituents. A paradigmatic example is the derivation of a Landau free-energy functional for magnetization starting from the lattice Ising Hamiltonian. Each coefficient in the Landau expansion can be written explicitly in terms of microscopic parameters, such as the exchange constant, temperature, and lattice dimensionality~\cite{Altland:2010aa}. Having access to both microscopic and effective levels of description allows one to combine analytical tractability with mechanistic insight.

Recently, the Landau formalism has been applied to describe the automatic perception of an interoceptive sensation~\cite{Valenzuela_Frontier2024}. 
In the model, the order parameter distinguishes between three minimal perceptual regimes: neutral, alert--protection, and trust--explore. These three regimes capture, in a minimal coarse-grained form, how a living organism may orient itself toward an inner sensation depending on its perceived significance for self-maintenance and action\cite{Jonasbook, Di-Paolo:2017aa}.
In the face of uncertainty, a living organism may adopt one of three states: an alert-protection state oriented toward adaptive survival, a trust-explore state facilitating expansion, and a neutral state representing the absence of motivational salience toward the inner sensation.
The Landau functional coefficients encode the influence of sensory context, embodied historical bias and socially mediated expert information. They determine which perceptual stance is stabilized. Within this model there is no universally preferred state: clinically, the critical issue is whether the perceptual stance remains rigid across varying socio-enviromental contexts. This phenomenological approach captures key features of chronic pain, including the emergence of hysteresis loops that can trap perception in a protective state. It also explains how perceptual thresholds shift with personal history, and how contradictory expert information can induce bio-psychosocial feedback loops\cite{Valenzuela_Frontier2024}.
This phenomenology aligns well with emerging research on chronic pain, which implicates large-scale brain networks (default mode, salience and attentional systems) and interprets their altered dynamics in terms of persistent and self-reinforcing evaluative patterns \cite{Barrett:2015aa, Sterlingbook}. Moreover, it naturally connects to enactive accounts of mental health as a graded property of organism-environment regulation across contexts\cite{Garcia:2025aa}, rather than a binary categorization into {\it adaptive} versus {\it maladaptive} patterns~\cite{Sterlingbook}.

In this context, a crucial missing element is a microscopic foundation that permits a rigorous derivation of the phenomenological Landau coefficients from an underlying microscopic framework. We identify the relevant microscopic level as the space of the internal evaluation of the organism with respect to the so-called vital norms: rules or patterns of interaction relied upon to maintain viability. The concept of vital norms is grounded in a longstanding tradition spanning philosophy of biology, phenomenology, and enactive cognitive science. Within this tradition, living beings are understood to generate their own norms related to metabolism, where the basic distinction between what supports and what threatens the continuity of life constitutes a primitive form of normativity and valuation\cite{Jonasbook}. The enactive approach further articulated a unified account of sense-making, according to which phenomena acquire meaning insofar as they relate, positively or negatively, to the organism's self-maintenance \cite{Di-Paolo:2017aa}. More recently, this framework has been expanded in the domain of mental health to emphasize the sensorimotor normativity of habits, integrating social norms into the dynamics of self-maintenance and thereby extending normativity beyond purely organic regulation \cite{Garcia:2025aa}.

\begin{figure}[tbh]
\centering
\includegraphics[width=\linewidth]{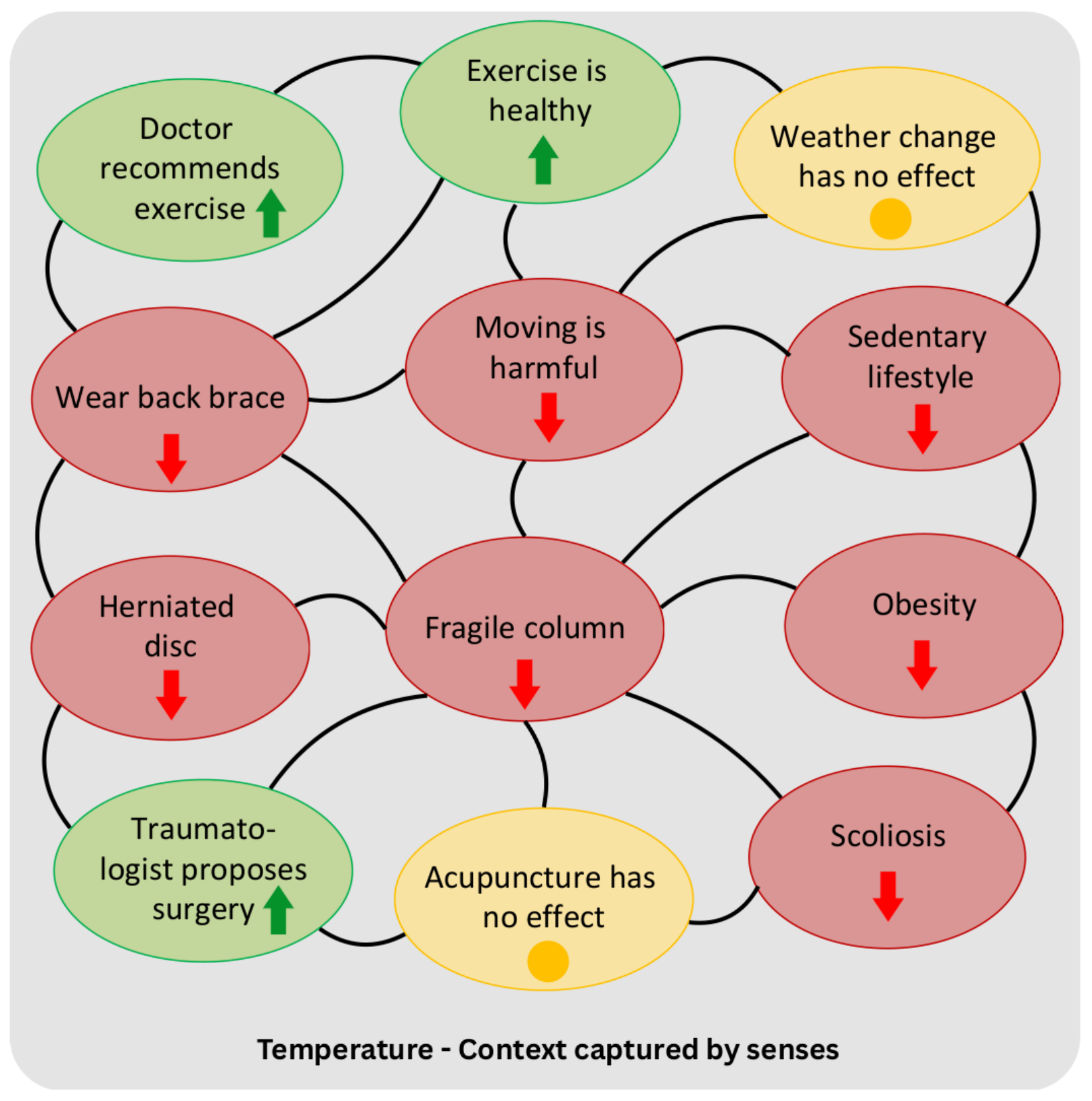}
\caption{
Example of Internal evaluation relative to chronic low back pain. Each site represents a vital norm and the local spin configuration encodes the organism's evaluation of that specific norm: green up arrow for positive (trustworthy), red down arrow for negative (threatening), and yellow circle for neutral (indeterminate). The links that connect different sites indicate that evaluations are not independent, but influence each other through an interaction structure. The vital-norm network is embedded in a broader contextual background (gray area), modeled by the temperature, $T$, that quantifies the contextual noise with respect to the sensation under study.}
\label{fig:Norms}
\end{figure}

Within this perspective, we develop a microscopic model for perception in which we define vital norms as judgments and habits about the inner sensation. The judgments and habits may reinforce one another and, in enactive language, can traverse the organic, sensorimotor, and social domains. We model vital norms as classical spin variables whose configuration encodes the organism's evaluation of a given norm: a positive spin ($+1$) represents a norm perceived as trustworthy, a negative spin ($-1$) denotes a norm viewed as threatening, and (when included) a zero spin ($0$) corresponds to an indeterminate or neutral evaluation. The ensemble of coupled spins defines a structured network of interdependent norms that shapes the organism's response to itself and its environment, see example in Fig.\ref{fig:Norms}. In this framework, perception of an inner sensation is not the result of any individual rule, but of the collective configuration of internal evaluations. This makes the Ising model and its generalizations a natural theoretical foundation for modeling experiential states. 

We analyze three progressively richer $\mathbb{Z}_2$-symmetric spin models. The first is the spin-$1/2$ Ising model, that only allows for binary evaluation of the vital norms. We will use it as baseline. We then extend the analysis to the spin-$1$ model by including a neutral configuration among the allowed ones. Finally, we examine the Blume-Capel (BC) model~\cite{Blume_PR1966, Capel_Physica1966}, where a tunable energy parameter $\Delta$ modulates the weight of the spin neutral state in the ensemble. Each model admits a derivation of an effective Landau-type theory via a Hubbard-Stratonovich transformation followed by a saddle-point expansion, yielding closed-form expressions for the Landau coefficients\cite{Altland:2010aa, NegeleOrland1998}.

These effective theories reveal how the availability and energetic tuning of a neutral evaluative state shapes the global perceptual behavior. The spin-$1$ model, compared to spin-$1/2$, exhibits a lower critical temperature and a smaller quartic coefficient, reflecting increased entropy and a flattened free-energy landscape due to the presence of an unpolarized state. In the BC model, the weight of neutrality is controlled by a tunable parameter $\Delta$. The model interpolates smoothly between spin-$1$ (at $\Delta = 0$) and spin-$1/2$ (as $\Delta \ll 0$), while for positive $\Delta$, the quartic coefficient becomes negative, generating a first-order transition and symmetry-preserving bistability. These effects provide a concrete microscopic basis for the multistability, threshold shifts, and hysteresis observed in the Landau model of perception.

Unlike Ref.~[\onlinecite{Valenzuela_Frontier2024}], where the Landau description is presented for a homogeneous perceptual order parameter, here we allow the emergent perception to vary across the network. This naturally introduces a stiffness parameter that governs the cost of local variation in the order parameter.
Cognitively, this can be interpreted as a proxy for mental rigidity or flexibility. Together with entropy, it modulates how sensitive the system is to context and how easily it transitions between perceptual regimes.\\

By integrating methods and perspectives from condensed matter field theory and cognitive science, this work grounds the phenomenological model in explicit microscopic structure and clarifies how perceptual evaluations, contextual modulation, and global perceptual organization emerge from basic spin interactions.
This integration not only advances fundamental cognitive understanding, but also departs from models focused solely on neural correlates or Bayesian inference by formalizing how internal evaluative structures generate macroscopic perceptual states. Our spin-based evaluative framework links microscopic evaluation processes with perceptual dynamics that may be relevant to clinical conditions. 
In this sense, the framework offers a useful theoretical language for discussing conditions such as chronic pain, anxiety, and related experiential disorders at a qualitative level.
This work further opens a path toward future generalizations involving asymmetric couplings, evolving evaluation patterns, and complex network architectures, key ingredients for modeling chronic and context-sensitive experiences.\\

The paper is organized as follows. Section~\ref{sec:Landau} reviews the phenomenological Landau model of perception and its key features. Section~\ref{sec:Models} presents the derivation of the effective action for the three symmetric spin models considered and compare the changes in the Landau coefficients. In Section~\ref{sec:Discussion}, we interpret these results in the context of perception of an inner sensation and outline the possible neurocognitive significance of key parameters. Section~\ref{sec:Outlook} outlines possible generalizations and future directions. Conclusions are summarized in Section~\ref{sec:Conclusions}. In Methods we provide full detail of the calculations.

\section{Landau model for the automatic perception of an inner sensation}
\label{sec:Landau}
The Landau model of perception of an inner sensation introduced in Ref.~[\onlinecite{Valenzuela_Frontier2024}] describes a semiconscious, automatic evaluation of a sensation, that might become a symptom, based on embodied norms, contextual information, and social cues. The scalar order parameter $\phi$ encodes the global perceptual stance: negative values correspond to protective or survival-oriented appraisals (alert-protection), positive values reflect vitality and openness (trust-explore), and $\phi = 0$ denotes a neutral perceptual state. These three regimes are introduced not as an ontological partition of interoceptive experience, but as a minimal coarse-grained description of automatic appraisal toward a sensation. In this sense, the model is formulated at the level of effective macrostates, broadly consistent with literature emphasizing the hedonic, salience-related, and action-oriented organization of interoceptive appraisal \cite{Craig2009, Barrett:2015aa, Strigo_2016, LFBarrettNatHumBeh17, Greenwood2025}.

The model adopts the structure of a standard Landau free-energy functional, expanded to fourth order:
\begin{equation}
F(\phi) = -h_{\text{ext}}\phi + \frac{a}{2}\phi^2 + \frac{h_{\text{int}}}{3}\phi^3 + \frac{b}{4} \phi^4.
\label{eq:phenom_landau}
\end{equation}
$F$ is the effective Landau free-energy functional of the theory:
it assigns relative probability (and hence stability) to alternative macroscopic perceptual stances, $P(\phi)\propto e^{-F(\phi)}$. The minimization of $F$ is therefore a stability principle: local minima of $F(\phi)$ identify the most likely and most persistent perceptual evaluations, where automatic attention is placed. Because this stability structure formalizes which perceptual stance becomes favored under given internal and contextual conditions, in the enactive interpretation adopted here, it can be understood as a mathematical formulation of {\it sense-making}.
Each coefficient encodes a distinct source of influence on the perceptual evaluation; changing its value reshapes the functional and thereby shifts which minima exist and which perceptual stance is stabilized: 
\begin{itemize}
\item $h_{\text{ext}}$: an {\it external} bias, such as socially mediated cues or expert information that directly polarize the macroscopic perceptual stance. 
\item $a$: temperature-dependent quadratic coefficient controlling the stability of the high-temperature neutral state $\phi=0$: near the transition temperature $T_0$, its vanishing signals that the system is becoming increasingly susceptible to polarization. Formally $a$ is proportional to the inverse susceptibility of the order parameter\cite{Altland:2010aa}: $a(T)\ra 0$ signals the divergence of the susceptibility and the onset of the instability toward a finite $\phi$. Near $T_0$, we can write $a(T)=a_0(T-T_0)$, where $a_0>0$ is a finite, model-dependent prefactor that sets the scale (amplitude) of the susceptibility close to $T_0$.
\item $h_{\text{int}}$: an {\it internal} bias, history-dependent and accumulated through prior experience associated with the sensation. The presence of a cubic term in Eq.(\ref{eq:phenom_landau}) introduces asymmetry: depending on the sign and magnitude of $h_{\text{int}}$, the potential becomes skewed, favoring one perceptual state over another.
\item $b$: quartic coefficient that shapes the curvature of the functional around the minima and stabilizes the overall potential (requiring $b>0$).
\end{itemize}

While $T$ plays the physical role of temperature, in our framework it parametrizes the contextual sensory evidence, specifically how informative the overall context is with respect to the specific sensation under study: high $T$ corresponds to a noisy/weakly informative background (large non-informative context relative to sensation-relevant evidence), whereas low $T$ corresponds to a context that is largely informative about the sensation; $T$ is therefore not a proxy for the total amount of sensory input, but a measure of how much informative the sensory input is in relation to the sensation. 
In this interpretation, high $T$ favors a neutral, non-polarized appraisal, i.e. a single minimum at $\phi=0$. As $T$ is reduced, neutrality destabilizes at the threshold $T_0$ and stable polarized minima emerge, so that below $T_0$ the system can stabilize either in an alert--protection ($\phi<0$) or in a trust--exploration appraisal ($\phi>0$), depending on internal and external biases. 
Clinically, $T_0$ can be read as a threshold where relatively small changes in context, such as a brief reassurance or a sudden loss of support, can trigger large shifts in the stabilized perceptual stance.

Within the Landau-type description, the discrete regimes correspond to distinct minima of $F(\phi)$, i.e. to alternative stabilized perceptual stances for given internal, external, and contextual conditions. At the same time, the notion of a {\it graded} continuum is captured by how these conditions vary continuously across individuals and contexts, and by how the corresponding features of the functional evolve continuously: the depth of minima, the height of the barriers between them, the width of metastability ranges, and the presence/strength of hysteresis. This provides a concrete Landau-language implementation of the enactive idea of graded, context-dependent regulation\cite{Di-Paolo:2017aa, Garcia:2025aa}: discrete perceptual stances are possible, while the degree of persistence versus reconfiguration is controlled by the evolving barrier structure as conditions vary.

In the Landau description \cite{Valenzuela_Frontier2024} the coefficients of the potential are introduced as free phenomenological parameters, in the present work instead we analytically derive them in terms of the microscopic parameters of the underlying spin models.

\begin{figure*}[tbh]
\centering
\includegraphics[width=0.98\textwidth]{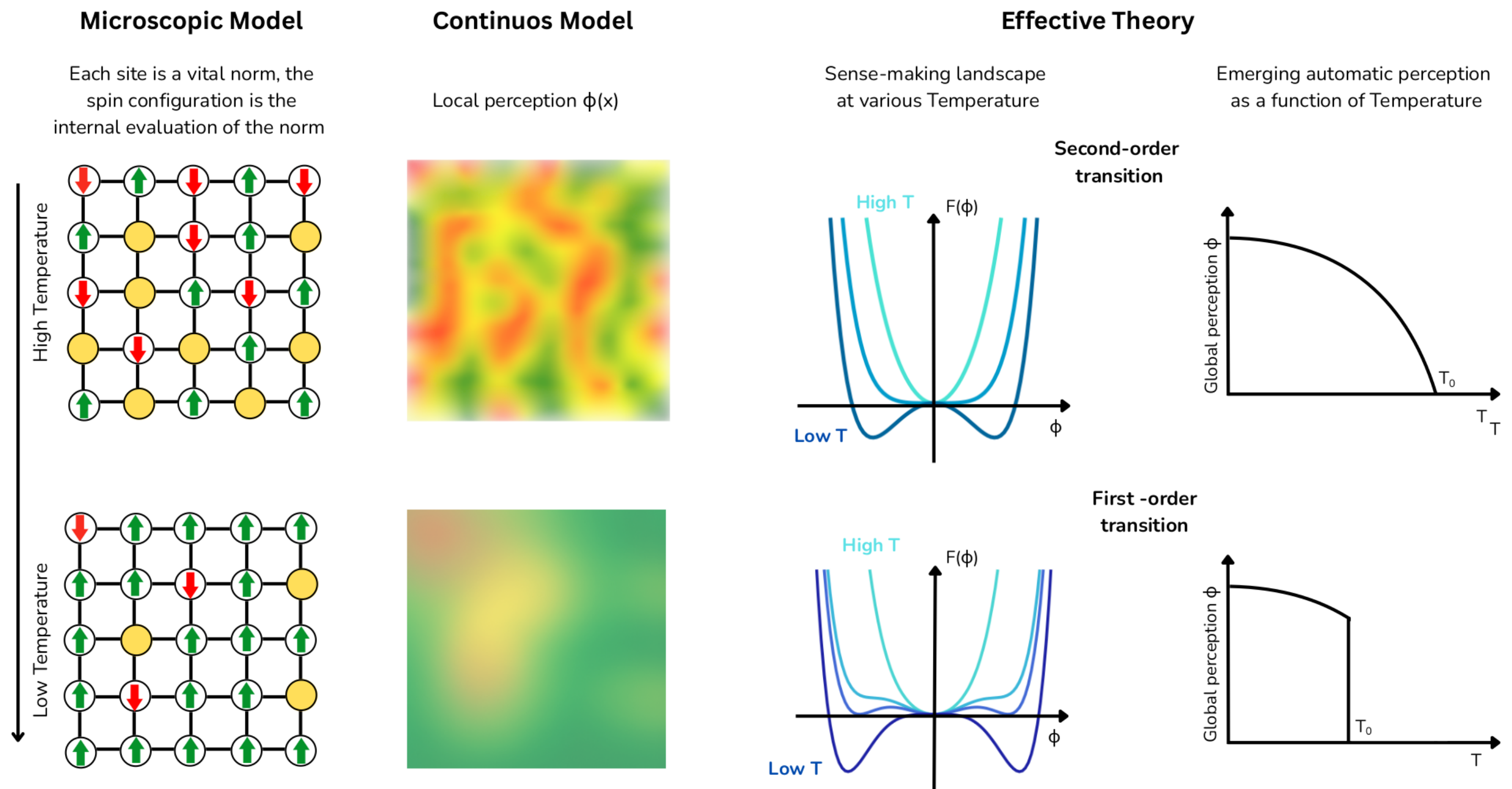}
\caption{\textbf{Emergent perception from internal evaluation.}
\textbf{Microscopic model:} each lattice site hosts a spin that represents the organism's local evaluation of a vital norm: green (up) for positive (trustworthy), red (down) for negative (threatening), and yellow (circle) for neutral (indeterminate). At high temperature the system is disordered, with coexisting positive/neutral/negative evaluations; at low temperature it orders toward a polarized configuration (positive in this example).
\textbf{Continuous model:} Corresponding coarse-grained perceptual field $\phi(x)$ produced by the local evaluations at high temperature (top) and low temperature (bottom). The cost of spatial fluctuations of $\phi(x)$, represented here by the shaded colors, is controlled by the stiffness parameter $\kappa$. The spatial profile of $\phi(x)$ shown here is intended only to illustrate a generic inhomogeneous coarse-grained field, rather than the solution of a specific spatial problem.
{\bf Effective Theory}: temperature evolution of the sense-making $F(\phi)$ controlling the stability and transitions between perceptual states for two different cases: (above) a symmetric double-well functional corresponding to spontaneous second-order transition emerges at low temperature, and (below) a non-convex potential indicating a first-order (jump-like) transition is found at low $T$.
In the Blume-Capel model, such first-order behavior can occur for moderate positive $\Delta$, where the quartic coefficient becomes negative, $b(\Delta)<0$; in this regime stability is restored by retaining the sixth-order term in the $\phi$ expansion (see Methods C).}
\label{fig:schematic}
\end{figure*}

\section{Ising models for the automatic perception of an inner sensation}
\label{sec:Models}

We now turn to the microscopic spin framework conceptually introduced earlier (Fig.~\ref{fig:Norms}). Across all models the {\it sites} represent the same network of vital norms relevant to the appraisal of an inner sensation, while the {\it spin variable} on each site encodes the local evaluation of that norm. The models differ only in the allowed local states: in the spin-$1/2$ case $S_i=\pm1$, whereas in the spin-1 and Blume--Capel cases a third, neutral option $S_i=0$ is included. Our aim is to derive a coarse-grained effective theory from these microscopic interactions and connect it to the phenomenological Landau functional in Eq.~\eqref{eq:phenom_landau}.

We analyze the system in the absence of any external field $h_{\text{ext}}$ and focus on the spontaneous emergence of perception. We want to isolate the role of internal evaluative interactions in shaping the sense-making landscape $F(\phi)$.

Our analysis is restricted to $\mathbb{Z}_2$-symmetric models, meaning that no symmetry-breaking terms, whether explicit or arising from built-in microscopic biases, are included. The derived effective theory therefore contains only even-order terms in the order parameter $\phi$. While the phenomenological model discussed in Eq.~(\ref{eq:phenom_landau}) incorporates a cubic term, we focus here on symmetric interactions that can nevertheless generate bistability and hysteresis through other mechanisms. The inclusion of asymmetric couplings remains an important direction for future work.

For what concerns the lattice structure of the microscopic model, it is important to clarify that this is not a spatial lattice in the physical sense, but rather a {\it conceptual network} of internal evaluative vital norms. Each site corresponds to a vital norm, and the interaction matrix $J_{ij}$ specifies the strength and range of their coupling. The most natural modeling is not obvious, since it may reflect hierarchical organization, heterogeneous connectivity, and dynamical reconfiguration typical of cognitive processing, leading to a complex and possibly disordered interaction network. 
In this work we adopt a regular nearest-neighbor lattice as a minimal, analytically tractable setting that enables a transparent microscopic-to-macroscopic derivation of the effective Landau landscape and its coefficients. This simplified structure has coordination number $z=2d$, i.e. each norm interacts with $2d$ others, and the lattice spacing $\ell_0$ represents the logical proximity between norms. This regular structure defines the microscopic scale in the continuum limit.
Importantly, the derivation does not rely on the specific geometric choice of $J_{ij}$. Within the present Hubbard--Stratonovich analysis, any finite-range modification of the interaction structure changes the effective theory only quantitatively: it shifts the transition scale and renormalizes the Landau coefficients, while leaving the overall Landau form of the effective functional unchanged. This minimal choice for lattice geometry and interactions structure is not essential to the mechanism discussed here, but is adopted because it makes the coarse-grained derivation analytically transparent and the resulting coefficients easy to interpret. Further details and examples are given in Methods D.
A more realistic modeling of the topology of internal evaluations, potentially informed by cognitive architecture or neurobiological constraints, will be considered in the future.

For completeness, all the details concerning the general derivation of the coarse-grained effective action are discussed in Methods A-C. The schematic of the process that lead from the microscopic model to the Landau-type functional is sketched in Fig.\ref{fig:schematic}. Below, we summarize the main steps and results for each model. 

\subsection{Spin-$1/2$ Ising model}

We begin with the spin-$1/2$ Ising model, in which each site $i$ carries a two-state spin variable, $S_i=\pm1$, encoding the local evaluation associated with a given vital norm. $S_i=+1$ corresponds to a positive (trustful) evaluation and $S_i=-1$ to a negative (threat-related) one.

The Hamiltonian takes the standard ferromagnetic form,
\begin{equation}
H[\{S\}] = -\frac{J}{2} \sum_{\langle i,j \rangle} S_i S_j,
\label{eq:Ising-H}
\end{equation}
where $S_i$ encodes the evaluation of the $i$-th vital norm and $J$ is the nearest-neighbor coupling between evaluations on the vital-norm network. For $J>0$ the interaction favors alignment of neighboring evaluations.

To describe the system's statistical behavior, we introduce the partition function $Z$, which encodes the equilibrium probability distribution over spin configurations
\begin{equation}
\label{eq:Z}
Z = \sum_{\{S_i = \pm 1\}} \exp\left( \frac{\beta J}{2} \sum_{\langle i,j \rangle} S_i S_j \right), \quad \text{with} \quad \beta = \frac{1}{T}.
\end{equation}
All macroscopic properties of the system, such as the average perceptual stance or susceptibility to change, can be derived from $Z$. In particular, it serves as the starting point for defining a coarse-grained effective theory. The temperature $T$ weights the Boltzmann exponential factors in Eq. (\ref{eq:Z}). Consistently with the interpretation introduced in Sec.~\ref{sec:Landau}, $T$ is a proxy for contextual noise with respect to the sensation under study (i.e., how weakly informative the overall context is for that appraisal). In the high-$T$ regime thermal fluctuations dominate and the spin lattice does not develop a net polarization, hence no coherent macroscopic perceptual stance. As $T$ is reduced, alignment is progressively favored and a coherent macroscopic stance emerges. For notational simplicity we have set Boltzmann's constant $k_B=1$ throughout.

To derive a coarse-grained field theory, we apply a Hubbard-Stratonovich transformation to decouple the spin-spin interactions in Eq.~(\ref{eq:Z}) and introduce a continuous scalar field $\phi(x)$, interpreted as the local average perception across many coupled evaluations. Assuming slow spatial variation, we obtain a standard Landau--Ginzburg action
\begin{equation}
\mathcal{S}_{\text{eff}}[\phi] = \int d^d x \left[ \frac{a}{2} \phi(x)^2 + \frac{\kappa}{2} (\nabla \phi)^2 + \frac{b}{4} \phi(x)^4 \right], 
\label{eq:phi4-action}
\end{equation}
where $\phi$ corresponds to the continuous macroscopic perception introduced in the phenomenological model Eq.~(\ref{eq:phenom_landau}), while the coefficients are now derived in terms of the microscopic parameters as shown in Table~\ref{tab:model-comparison} (see Methods A for full derivation). 
As expected, the quadratic coefficient takes the form $a = a_0 (T - T_0)$. However, the effective action derivation allowed us to determine the explicit form of $a_0$ and $T_0$. The critical temperature $T_0$ scales with the coordination number $2d$ and the interaction strength $J$. Once $T$ drops below this threshold, the neutral state becomes unstable and spontaneous perceptual polarization emerges even in the absence of an external field. The same microscopic scales also control the quartic coefficient.

Allowing for slow spatial variations of the coarse-grained perceptual field $\phi(x)$ across the vital-norm network introduces a gradient term in Eq.~(\ref{eq:phi4-action}). The stiffness coefficient, which reflects the homogeneity of the perceptual field, is given by $\kappa = J \ell_0^2 /T$, where $\ell_0$ is the lattice spacing. Although not discussed in the phenomenological model\cite{Valenzuela_Frontier2024}, $\kappa$ plays an important role in determining how perception propagates across the evaluative system and, as we will discuss in the next section, may be interpreted as a measure of mental rigidity or flexibility within that network.

Because the model is fully $\mathbb{Z}_2$-symmetric, the effective action contains only even powers of $\phi$ and, for $T<T_0$, develops two degenerate minima at $\pm\phi_0$, corresponding to alternative perceptual regimes. This model provides a reference point against which we will evaluate the influence of additional spin states or tunable parameters in more elaborate models.

\subsection{Spin-$1$ Ising model}

We now consider the spin-$1$ Ising model on the same vital-norm network introduced above, but with an enlarged local state space:
each site $i$ carries a spin variable $S_i \in \{-1, 0, +1\}$. The values $\pm 1$ represent positive and negative evaluations of a vital norm, as in the spin-$1/2$ case, while $S_i = 0$ corresponds to a neutral evaluation, i.e. the organism registers a norm as present but not salient or currently relevant. 
The inclusion of this neutral state adds a qualitatively new degree of freedom to the model, allowing the system to withhold judgment in ambiguous conditions and thereby modulating the robustness of ordered perceptual states.

The Hamiltonian is the same of Eq.(\ref{eq:Ising-H}) and the partition function is defined as in Eq.~\eqref{eq:Z}, but the sum now runs over $S_i \in \{-1, 0, +1\}$. Following the same steps as in the spin-$1/2$ case, we derive a coarse-grained field theory. The presence of the neutral state modifies the local spin distribution and its cumulants, leading to a different set of coefficients in the resulting effective potential. Assuming a slowly varying field $\phi(x)$ and expanding the partition function, we again obtain an effective action of the form given in Eq.~(\ref{eq:phi4-action}), but with renormalized coefficients reflecting the influence of the additional spin state. Results shown in Table~\ref{tab:model-comparison}.

The mass term again takes the form $a = a_0(T - T_0)$ but the critical temperature is reduced compared to the spin-$1/2$ case.
The stiffness coefficient retains its previous form, while the quartic coupling becomes smaller as result of the possible neutral configuration of the spins.

The lower value of $T_0$ reflects the fact that the availability of a neutral option ($S_i=0$) stabilizes the non-polarized regime and reduces the system's tendency to polarize. In perceptual terms, the presence of a noncommittal evaluative option makes it less likely for the system to settle into a coherent stance unless the context is strongly structured. The reduction in $b$ indicates a flatter effective potential, with the added entropy from the $S=0$ state softening the transition and enhancing sensitivity to fluctuations.

The spin-$1$ model thus extends the binary perception framework by introducing a neutral configuration, which weakens the tendency to order compared to the spin-$1/2$ case.

\begin{table*}[tbh]
\begin{center}
\begin{tabular}{l@{\hskip 30pt}c@{\hskip 30pt}c@{\hskip 30pt}c}
\hline
\textbf{Model} & \textbf{$a_0$} & \textbf{$T_0$} & \textbf{$b$} \\
\hline
\\
Spin-1/2 Ising &
$\displaystyle \frac{2dJ}{T^2}$ &
$\displaystyle 2dJ$ &
$\displaystyle \frac{1}{3} \left(\frac{2dJ}{T}\right)^4$ \\
\\
Spin-1 Ising &
$\displaystyle \frac{2dJ}{T^2}$ &
$\displaystyle \frac{4dJ}{3}$ &
$\displaystyle \frac{1}{9} \left(\frac{2dJ}{T}\right)^4$ \\
\\
Blume--Capel &
$\displaystyle \frac{2dJ}{T^2} \left(1 - \frac{\Delta}{4dJ}\right) $ &
$\displaystyle \frac{4dJ}{3}\left(1 - \frac{\Delta}{4dJ}\right)$ &
$\displaystyle \frac{1}{9}\left(\frac{2dJ}{T}\right)^4\left(1 - \frac{\Delta}{T}\right)$ \\
\\
\hline
\end{tabular}
\caption{{\bf Landau parameters for Ising-like models.} $T_0$ is the transition scale, defined by $a(T_0)=0$), $a_0$ is the linear coefficient in the critical expansion $a(T)\simeq a_0\,(T-T_0)$ that sets how fast the quadratic term grows away from criticality, and $b$ is the quartic coefficient of the Landau landscape. The expressions are exact for the spin-$1/2$ and spin-$1$ models, while for the Blume--Capel (BC) model we report the leading (linear) corrections in $\Delta$ around $\Delta=0$ (see Methods for the full derivation). Comparing the models highlights the impact of allowing a neutral microscopic configuration ($S_i=0$). Relative to spin-$1/2$, the spin-$1$ model is less prone to order (smaller $T_0$) and displays a flatter Landau landscape (smaller $b$) more sensitive to fluctuations. In the BC model, $\Delta$ penalizes ($\Delta>0$) or favors ($\Delta<0$) the polarized states $S_i=\pm1$, thereby tuning the weight of the neutral state. To leading order, $\Delta>0$ further reduces $T_0$ and $b$, and also lowers $a_0$ (so $a(T)$ increases more slowly above $T_0$), whereas $\Delta<0$ produces the opposite trend, strengthening the tendency to order. For sufficiently positive $\Delta$, $b(\Delta)$ becomes negative, signaling a discontinuous (first-order) regime that is stabilized by the higher-order $\phi^6$ term (see Methods section).}
\label{tab:model-comparison}
\end{center}
\end{table*}

\subsection{Blume-Capel model}

We now consider the BC model\cite{Blume_PR1966, Capel_Physica1966}, a natural extension of the spin-1 Ising model that introduces a tunable energetic bias toward or against neutral evaluation. This extension captures an important cognitive feature of perceptual processing, i.e. the ability of the organism to modulate salience or suppress the relevance of certain sensations. In perceptual and attentional terms, this can be interpreted as the capacity to down-regulate or up-regulate the tendency to remain undecided or disengaged with respect to a sensation.

Each spin variable again takes values $S_i \in \{-1, 0, +1\}$, with $\pm 1$ representing polarized evaluative states and $0$ indicating a neutral or non-salient evaluation. The Hamiltonian is modified to include a single-site potential:
\begin{equation}
H[\{S\}] = -\frac{J}{2} \sum_{\langle i,j \rangle} S_i S_j + \Delta \sum_i S_i^2.
\label{eq:BC-H}
\end{equation}
The parameter $\Delta$ controls the energetic cost of the polarized states. When $\Delta > 0$, the $S_i = \pm 1$ configurations are penalized, favoring the neutral state, $S_i = 0$. Conversely, when $\Delta < 0$, the system energetically favors polarized evaluations, suppressing neutrality. 
In the present framework, positive $\Delta$ therefore stabilizes a non-committal evaluative stance, whereas negative $\Delta$ promotes stronger evaluative commitment. We discuss possible neurocognitive interpretations of this tendency in more detail in Sec.~\ref{sec:Discussion}.
The partition function becomes
\begin{equation}
Z = \sum_{\{S_i\}} \exp\left( \frac{\beta J }{2} \sum_{\langle i,j \rangle}  S_i S_j - \beta \Delta \sum_i S_i^2 \right),
\end{equation}
and we proceed as before by applying a Hubbard-Stratonovich transformation.

The resulting effective action again takes the form of Eq.~\eqref{eq:phi4-action}, but with $\Delta$-dependent coefficients. Full details of the derivation  can be found in Methods C. In Table~\ref{tab:model-comparison} we report the expressions at the lower order of correction in $\Delta$. As expected, at $\Delta=0$ the BC coefficients reduce continuously to the spin-$1$ values. 

The mass term remains $a=a_0(T-T_0)$, but the transition scale $T_0$ and the slope $a_0$ become $\Delta$-dependent, for small $|\Delta|$ their leading corrections are linear. For $\Delta<0$, $T_0$ increases monotonically, reflecting the growing suppression of the neutral configuration: $S=0$ is disfavored and the system progressively approaches a more binary (spin-$1/2$-like) behavior, ordering more readily. Conversely, for $\Delta>0$ the neutral state is favored and $T_0$ is reduced, so that a neutral perceptual stance $\phi$ persists down to lower contextual noise (i.e., a more informative context is required to destabilize the unpolarized phase). 
The prefactor $a_0$ changes in the same direction and controls how quickly the system moves away from the critical region above $T_0$. For $\Delta>0$, $a_0$ is reduced, so that $a(T)\simeq a_0(T-T_0)$ increases more slowly with $T-T_0$, implying a weaker curvature of the $\phi=0$ minimum just above $T_0$. For $\Delta<0$, instead, $a_0$ increases and $a(T)$ grows more rapidly, so that the system departs more sharply from criticality, i.e. the critical regime is crossed over a narrower temperature range.

The stiffness coefficient $\kappa$ remains unchanged, while the quartic coefficient $b$ becomes $\Delta$-dependent. The leading small-$|\Delta|$ expressions reported in Table~\ref{tab:model-comparison} already show the qualitative trend: for $\Delta<0$ the quartic coefficient of the effective potential is enhanced resulting in a stiffer landscape with more strongly confining minima, for $\Delta>0$ instead it is reduced resulting in a flatter landscape. This decrease of $b$ for $\Delta>0$ anticipates the possible onset of a regime where $b$ vanishes and the quartic theory ceases to be stable. To make the emergence of the first-order regime transparent, and to verify that it is not an artifact of the small-$|\Delta|$ expansion, it is useful to consider the full expression (Methods~C), which can be factorized as
\begin{equation}
b(\Delta)=\mathcal{B}(\Delta/T)\,\big(4-e^{\Delta/T}\big),
\label{eq:bsimple}
\end{equation}
where $\mathcal{B}(\Delta/T)$ is strictly positive. Therefore the sign of $b(\Delta)$ is controlled entirely by the factor in parenthesis that changes sign at a value $\Delta^*= T \ln 4$. For $\Delta>\Delta^*$, the effective potential becomes non-convex and the system undergoes a first-order transition, characterized by bistability and hysteresis \cite{ChaikinLubensky, StrukovLevanyuk, Binder:1987aa, Wang2010}. In this regime stability is restored by retaining the next non-vanishing even term in the Landau expansion (the $\phi^6$ term), as discussed in Methods~C.

In conclusion, the BC model reduces to the unbiased spin-$1$ case at $\Delta=0$. For $\Delta<0$, the neutral configuration is increasingly suppressed and the coefficients continuously approach the spin-$1/2$ limit, corresponding to a steeper, purely convex effective potential and more robust polarization. For positive $\Delta$ the system displays opposite behavior and for $\Delta>\Delta^*$ undergoes a first-order transition with bistability and hysteresis even in the absence of explicit asymmetry. In this way the model reproduces a key qualitative feature of the phenomenological Landau theory, namely the emergence of multiple locally stable perceptual stances from a symmetric Hamiltonian. Overall, tuning $\Delta$ modulates the tendency toward neutrality or polarization and thereby reshapes the structure of the sense-making landscape. In particular, the appearance of multiple minima and hysteresis loops provides a concrete microscopic route to persistent perceptual patterns, such as those discussed in the context of chronic symptom perception.

\section{Discussion}
\label{sec:Discussion}

\subsection{Implications of the Model and Neurocognitive Interpretation}

We treat perception of an interoceptive sensation as an emergent property of a deeply structured evaluative process, shaped by embodied norms, significant context, and internal organization. The Landau model introduced in Ref.~[\onlinecite{Valenzuela_Frontier2024}] captured this perspective phenomenologically, showing how perception of an inner sensation can become polarized, modulated, or dynamically trapped through the interplay of internal structure and external influences.

Here, we have provided a microscopic foundation for that model, deriving effective Landau-like theories from a class of $\mathbb{Z}_2$-symmetric spin models. Starting from increasingly expressive Hamiltonians, ranging from the spin-$1/2$ Ising model to the BC extension, we have shown how key features, such as bistability, perceptual thresholds, rigidity, and context-sensitive transitions can emerge from the underlying structure of internal evaluative interactions.

In this framework the spin variables represent internal evaluations of vital norms that may be distributed across cognitive systems. Their interactions encode how the organism integrates embodied norms to generate a global perceptual configuration. 
The order parameter $\phi$ captures this emergent evaluation as a macroscopic measure of the system's collective stance. As such, $\phi$ transcends the brain: is not confined to neural variables alone, but reflects the configuration of a distributed evaluative system that spans brain, body, and environment. This interpretation aligns with perspectives from embodied and extended cognition, where sensorimotor, or contextual couplings all contribute to perceptual organization~\cite{EmbodiedMindVarela92, Fristonautopoiesis18, LFBarrettNatHumBeh17}. Tuning the model's parameters allows exploration of context sensitivity, perceptual rigidity, and the effects of embodied vital norms. In this setting, temperature $T$ is a proxy for contextual noise with respect to the sensation under study, i.e. how weakly informative the overall context captured by the senses is for that appraisal. The distance to the threshold, $(T-T_0)$, quantifies whether the context is sufficiently informative to shift the system away from a neutral stance and toward a polarized perceptual evaluation, and it controls the instability via $a\sim (T-T_0)$, which vanishes as $(T-T_0)\to 0$.

The parameter $\Delta$, modulates the energetic cost of neutrality, which affects the perceptual threshold $T_0(\Delta)$ and thus tunes the system's sensitivity to salience.
A positive $\Delta>0$ favors neutral configurations at the microscopic level and lowers the critical threshold $T_0$. As a result, a neutral (non-polarized) perceptual stance $\phi\simeq 0$ may arise not only as a high-noise effect (large-$T$ behavior), but also as an explicit energetic bias toward neutrality that stabilizes $\phi\simeq 0$ even when the sensory context is highly informative (low $T$). 
In cognitive terms, this may correspond to a reduced tendency for the sensation to acquire salience and drive a polarized appraisal, potentially reflecting attentional disengagement, experiential suspension, or perceptual flatness. This offers a principled way to link microscopic model structure to regulatory phenomena involving large-scale networks such as the insula cortex~\cite{Craig2009}, the salience system~\cite{Seeley2007}, or the default mode network~\cite{Barrett:2015aa}.
Conversely, when $\Delta < 0$, neutral states are penalized and the system more readily adopts polarized configurations. This results in a higher critical temperature $T_0$ and may reflect pathological forms of hyper-attunement to sensory input. Clinically, such dynamics may underlie phenomena observed across several disorders. For example, allodynia or hyperalgesia in chronic pain, where non-noxious or mildly noxious stimuli are perceived as intensely painful or salient, may reflect a heightened tendency toward perceptual polarization\cite{Kaplan:2024aa}.  
In anxiety disorders, the organism often responds with alarm in a way that is disproportionate to the context, reflecting an underlying evaluative process\cite{Craske:2017aa}. Similarly, in functional disorders, where symptoms are experienced despite the absence of structural lesions that justify them, the organism exhibits a disproportionate evaluative bias toward the symptom\cite{Fiorio:2022aa}.
The dual interpretation of $\Delta$ for positive and negative values, offers a bridge between model parameters and testable neurophysiological biomarkers and behavioral observations. For instance, interventions such as embodied mindfulness, known to enhance habits, judgement and interoceptive awareness of the semiconscious perception and promote nonjudgmental processing of internal signals\cite{Keng2011, Greenwood2025}, may act by dynamically modulating $\Delta$, thereby shifting the system toward greater neutrality or reduced evaluative reactivity.

A central result of our analysis is that the inclusion of a neutral evaluative state, $S_i = 0$, in the spin-$1$ and BC models qualitatively transforms the system's behavior. The presence of a non-polarized option increases the entropy of the system and lowers both the critical temperature and the quartic coefficient. As a result, the effective potential becomes flatter and more sensitive to contextual information, increasing the system's responsiveness and flexibility. From a neurocognitive standpoint, this reflects the capacity to withhold judgment, disengage from immediate response, or tolerate uncertainty. Such neutrality may be adaptive in environments with ambiguous or conflicting signals, allowing the organism to delay action until more meaningful information is available. Physiologically, it may involve regulatory mechanisms that inhibit perceptual polarization, such as increased default mode activity~\cite{Barrett:2015aa}, insular modulation of interoceptive salience~\cite{Craig2009}, or reduced salience network reactivity~\cite{Seeley2007}. Clinically, persistent occupation of neutral or disengaged states may relate to phenomena such as chronic indecision, emotional numbing, or experiential dissociation~\cite{Posner:2005aa}, which are common features of depression, and chronic apathy. In this light, allowing the local state $S_i=0$ provides a microscopic route to a macroscopic near-neutral stance $\phi\simeq 0$, which can act either as an adaptive buffer against premature polarization or, if overly persistent, as a source of experiential stagnation.

A second key result concerns the emergence of multistability and hysteresis in fully symmetric models. Although the microscopic Hamiltonians we study contain no explicit symmetry-breaking terms, the BC model exhibits a transition from second- to first-order behavior as the neutral state becomes energetically favored. When $\Delta$ is sufficiently positive, the quartic coefficient $b(\Delta)$ becomes negative, producing a non-convex potential with coexisting minima. This demonstrates that bistability and hysteresis loops, hallmarks of the phenomenological model, can arise purely from entropy-driven mechanisms and internal tuning, without requiring cubic asymmetries or external bias. From a cognitive standpoint, this connects directly to the previous discussion. When neutrality becomes structurally dominant, the system may become trapped in ambiguous or disengaged configurations. The resulting multistability may, if permanent, underlie pathological forms of experiential suspension, such as indecision, perceptual blunting, or chronic uncertainty.

While the phenomenological model of Ref.~[\onlinecite{Valenzuela_Frontier2024}] focused on the spatially homogeneous Landau functional, our microscopic derivation allows for spatial variations of $\phi(x)$ across the vital-norm network. This yields a gradient term and the associated stiffness $\kappa$. $\kappa$ quantifies the cost of local variation in the order parameter $\phi$, and thus controls the coherence of the perceptual field. High stiffness enforces strong internal alignment, potentially corresponding to cognitive rigidity or global overgeneralization. Low stiffness, by contrast, permits local differentiation, context-specific responses, or compartmentalized evaluations. Since $\kappa$ increases with interaction strength $J$ and lattice spacing $\ell_0$, while decreasing with temperature $T$, it captures a trade-off between internal connectivity and responsiveness to external sensory information. High $J$ (or large $\ell_0$) favors globally aligned perceptual configurations, whereas higher $T$ softens the perception, promoting  fragmented evaluations.
While $\kappa$ is a standard parameter in field-theoretic models, we propose that it may serve here as a structural correlate of perceptual coherence, modulating the balance between global alignment and local flexibility in evaluative dynamics.
This interpretation resonates with findings in clinical neuroscience linking cognitive rigidity to altered network dynamics in mood and anxiety disorders~\cite{Etkin2015}, and to default mode hyperconnectivity observed in depression and related conditions~\cite{WhitfieldGabrieli2012}. It is also consistent with work suggesting that reduced network flexibility underlies impaired context switching and evaluative adaptation in neurodivergent populations~\cite{Schlosser2010}. These perspectives suggest that $\kappa$ may provide a theoretically grounded proxy for a dimension of cognitive style, shaping resilience, vulnerability, or the persistence of suboptimal perceptual states\cite{Garcia:2025aa}.

\subsection{Relation to Other Modeling Frameworks}

To further situate our framework within the broader landscape of cognitive and psychological modeling, we now compare it with several existing models that share formal similarities or address overlapping themes in perception and cognition.
While each originates from a different theoretical motivation, they illustrate how similar mathematical tools or conceptual motifs are used across domains, and help clarify the distinct perspective developed in our work.\\

Ising-like models have been widely used in theoretical neuroscience, particularly in modeling large-scale brain dynamics and criticality. Neural spin models can reproduce key statistical properties of brain activity and are often analyzed near critical points, where enhanced sensitivity and dynamic range are proposed to support cognitive flexibility~\cite{diSanto2018, Ruffini2023}. In this tradition, mental pathologies are sometimes linked to departures from criticality~\cite{Munoz2018}, associated with dynamical trapping, reduced responsiveness, or impaired adaptability.
In this study we explore how analogous features arise not from external tuning toward criticality (as it would appear in the presence of odd-order terms in the effective action\cite{Valenzuela_Frontier2024}), but from internal entropy-driven mechanisms in which the system undergoes a first-order transition, originated by dominance of neutral states, producing multistability and hysteresis. These features may correspond to perceptual uncertainty or experiential stagnation, particularly in cases where disengagement or indecision becomes structurally stabilized. 

While both frameworks employ spin-like models, their building blocks differ fundamentally. In traditional neural models, spins often encode neuron-like activation states, typically aligned with anatomically localized computations. In contrast, our model interprets spins as functional units of internal evaluation that do not correspond to individual neurons or specific brain regions. These evaluative units emerge from
vital norms that can transverse organic, sensory-motor and social domains. This abstraction allows us to move beyond strictly neural interpretations and to model evaluative coherence at the level of the whole organism. Accordingly, our approach is not brain-centered; it emphasizes the systemic evaluative structure of the agent and its embedding in social and environmental context.
While traditional neural models often struggle to incorporate embodiment and sociocultural modulation in a principled way~\cite{McGann2020}, our spin variables explicitly represent context-sensitive norms. This enables us to model how internal evaluations and external signals jointly shape the perception of an interoceptive sensation. Bridging the explanatory gap between the physiological neural network models and the experiential structure formalized in our work constitutes part of the hard problem of consciousness~\cite{Chalmers1995-CHAFUT}, and lies beyond the scope of this study. However, it is noteworthy that both domains, physiological and experiential, can be described using a shared phase transition framework.

A second model we wish to contrast with is an inference-based account of perception combined with a neural network theory close to a phase transition, the REBUS model~\cite{Carhart-Harris:2019aa, Carhart-Harris_2023}. Both frameworks share the broader idea that interoceptive perception is not a passive readout of bodily signals, but depends on context-sensitive internal organization. In the REBUS model the brain is modeled as probabilistic inference, where beliefs are updated to minimize prediction error, weighted by expected precision. By contrast, our model treats perception as primary and evaluative, not as a byproduct of belief optimization, but as an emergent configuration of embodied norms embedded in significant contextual constraints to survive and expand.
This divergence is especially salient in how uncertainty is conceptualized. In REBUS, uncertainty reflects diminished confidence in a belief and is linked to increased plasticity or relaxed priors. In our model, by contrast, neutrality is not simply low confidence, but a structurally stabilized state of disengagement, suppression, or open-ended tolerance. This framing allows for bistability and hysteresis to arise even in fully symmetric systems, without requiring built-in asymmetries or shifts in precision. In short, our model shifts the emphasis from inference to evaluation, and from belief optimization to perception of an interoceptive sensation as the dynamic expression of interdependent internal evaluations of vital norms in an increasing significant context when $T$ is approaching $T_0$.\\

Lastly, spin-based models have also been developed within psychology, particularly in network approaches to mental disorders, attitudes, and cognitive skills~\cite{Borsboom2017, Dalege2016,Savi2019, vanderMaas_preprint}. In many of these models, each spin represents a psychological unit, such as a symptom, belief, habit or skill, and interactions capture causal or correlational influences between them. While this framework is empirically productive, the meaning of spin variables and their interactions is typically tailored to specific psychological constructs, e.g. symptoms, attitudes, or skills, depending on the domain of application. In contrast, our model provides a unified microscopic account of health-related perception, where spins always represent the organism's local evaluations of vital norms embedded in socio-enviromental context, and the interaction matrix governs how these evaluations integrate into coherent perceptual states.

A second key difference lies in the interpretation of the Hamiltonian. In psychological network models its cognitive interpretation is often unspecified: the Hamiltonian mainly serves as a formal tool to enforce coherence or consistency, and is often interpreted in abstract terms such as cost, dissonance, or utility~\cite{vanderMaas_preprint}. By contrast, in our framework the effective energy functional has a concrete, cognitive grounded meaning in the statistical-mechanics sense: by assigning relative probability (and hence stability) to alternative macroscopic perceptual stances, it defines the corresponding {\it sense-making} landscape. In this way, it directly reflects the internal evaluative architecture and shapes the emergence of perception through embodied, context-sensitive interactions. This enables a more principled connection between microscopic evaluative states and the macroscopic emergence of perception of an interoceptive sensation.
 
The role of temperature also differs. In psychological applications, inverse temperature, $\beta=1/T$, is often interpreted as a proxy for attention or motivational engagement~\cite{vanderMaas_preprint}. In our model, by contrast, $T$ is a proxy for contextual noise with respect to the sensation under study, i.e. how weakly informative the overall sensory context is for that appraisal. High $T$ corresponds to a noisy background about the sensation, which favors neutral or unpolarized perceptual states. Low $T$ corresponds to a more informative context, which promotes the stabilization of a specific perceptual evaluation of an inner sensation. These definitions are closely related, since both attentional capacity and contextual noise can affect the effective discriminability of sensation-relevant information. It is worth noticing, however, that whereas in those approaches attention enters as an input parameter modulating cognitive focus~\cite{vanderMaas_preprint}, in our framework it appears instead as an emergent macroscopic quantity associated with the global minimum~\cite{Valenzuela_Frontier2024}.

Despite these conceptual differences, formal similarities remain. For instance, the inclusion of a neutral ($S=0$) state in spin models has been shown to increase entropy and support multistability \cite{vanderMaas_preprint}. This reinforces the broader insight that perception and cognition can be fruitfully understood as the macroscopic consequences of constrained spin systems, though the conceptual cognitive frameworks are different.

\section{Outlook and Future Directions}
\label{sec:Outlook}

This work provides a link between microscopic spin models and the phenomenological model of perception\cite{Valenzuela_Frontier2024}. A general result is how Landau parameters are related to Ising parameters which can be useful for other proposals in cognitive Ising modeling. The framework is flexible and analytically tractable, and opens the way to natural extensions. At the same time, the present formulation involves several simplifying assumptions, which also indicate its main limitations and the most natural directions for extension.

To make the model more realistic, several generalizations are possible. 
A first limitation of the present formulation is the use of discrete scalar spins, with values $-1$, $0$, and $+1$ encoding evaluative states. Real internal judgments are typically graded rather than discrete, varying in strength across a continuous spectrum. A vector-valued model, such as the classical Heisenberg model, could therefore capture more nuanced evaluative states, where both the strength and the dimensional orientation of internal judgments are represented.

Beyond the structure of individual evaluative states, the symmetry of their interactions is also a key factor.
Many phenomena associated with persistent symptoms involve history-dependent biases, modeled phenomenologically by a cubic term in the Landau potential. 
While such terms do not arise in the symmetric models treated here, introducing microscopic symmetry-breaking mechanisms, such as asymmetric couplings, or three-spin interactions, could generate odd-order terms in the effective action, offering a more complete account of how embodied history shapes the sense-making landscape.

The structure of the interaction network also warrants further exploration. Our models use a square regular lattice, for analytical tractability, but real-world evaluative systems are likely modular, hierarchical, or dynamically reconfigurable. Graph-based, connectome-informed, or multilayer topologies could better capture the distributed organization of cognitive-perceptual systems and account for compartmentalized, conflicting, or selectively activated evaluations. A concrete example for such extensions is provided by structural connectomes used in network neuroscience, e.g. the Sporns connectome\cite{Sporns_AnnRevPsyco2016}, which has been widely employed as a mesoscopic graph on which to test statistical-physics dynamics.

Sensory context could also be refined. In our framework, temperature $T$ encodes the contextual noise respect to the sensation as a global parameter. A site-dependent $T_i$ would introduce spatial heterogeneity in the sensory information flow, allowing the model to represent selectively informative (or noisy) contexts and localized disruptions in appraisal.

Another limitation of the present formulation is its static character. The current models can be extended to include time-dependent changes. Learning, habituation, or therapeutic transformation could be modeled as slow, history-dependent shifts in parameters such as $\Delta$, $J$, or $T$, allowing the system to traverse different regions of the sense-making landscape over time as performed in similar models\cite{Carhart-Harris:2019aa,Aguilera:2018ab}. 
More structured forms of time evolution could also be formalized using multicomponent or hierarchical field theories, where each field governs a distinct layer of evaluative processing. For example, one could model automatic perception and conscious framing as two interacting fields, coupled through cross-terms in the effective action. This would enable feedback and modulation across layers without requiring full dynamical complexity.

Potts and Hopfield models are natural generalizations of the Ising framework, extending the allowed spin states and interaction structures. They could serve as microscopic foundations for more expressive Landau-type theories of perception, capturing additional internal states or richer connectivity. While widely used to model memory and retrieval~\cite{Kanter1987,Hopfield1982}, our focus is to explore how such models can organize perceptual evaluation, rather than information storage. This shift in emphasis, from memory to embodied valuation, may reveal new regimes of multistability or context-sensitive transitions in sense-making landscapes.

Altogether, these directions suggest a broader theoretical landscape in which microscopic models of valuation are linked to higher-order cognitive processes and embodied experience. Developing such multiscale frameworks, rooted in tractable spin-based architectures, may offer new insight into how perception becomes stabilized, disrupted, or reshaped in persistent conditions, and how it might change through neurobiological education, self-awareness techniques or intervention.

\section{Conclusions}
\label{sec:Conclusions}
We presented a principled microscopic framework that explains how large-scale perceptual features, such as contextual sensitivity, perceptual rigidity and hysteresis, can emerge from local evaluative processes grounded in embodied norms. By  explicitly deriving an effective theory from spin-based microscopic models, our approach bridges phenomenological and cognitive levels of analysis in a rigorous way. This connection offers mechanistic insights into persistent symptoms, such as chronic pain and mental disorders, and may inform future work on their modulation through neurobiological education, therapy, or self-awareness techniques.
In contrast to inference-based or neural-correlate models, our approach provides an understanding and generative account of perceptual organization. This work defines a flexible modeling strategy, that can be systematically extended to incorporate asymmetric interactions, dynamic adaptation, and realistic network topologies, laying the foundation for future research on the structure and plasticity of embodied experience.

\section*{ACKNOWLEDGMENTS}
The authors acknowledge Miguel A. Mu\~noz and Fabio Cecconi for critical reading of the manuscript and Brian Staroselsky for interesting discussions.

\section*{METHODS}
\subsection{Derivation of the Effective Theory for the Spin 1/2 Model}
\label{App:HS_Ising}

In this section we outline the derivation of the coarse-grained effective action used in the main text. The method is based on a standard Hubbard--Stratonovich transformation followed by a saddle-point expansion~\cite{NegeleOrland1998, Altland:2010aa}.\\

\noindent The starting point is the Hamiltonian
\begin{equation}
H[\{S_i\}] = -\sum_{\langle ij \rangle} J_{ij} S_i S_j,
\end{equation}
where $S_i = \pm 1$, and $J_{ij} > 0$ encodes ferromagnetic coupling. In the main text we simplify by assuming $J_{ij} = J$ for nearest neighbors, but here we keep the general notation to distinguish model-independent steps from lattice-specific ones.
The corresponding partition function reads
\begin{equation}
Z = \sum_{\{S_i\}} \exp\left( \sum_{\langle ij \rangle} \tilde{J}_{ij} S_i S_j \right),
\end{equation}
with $\tilde{J}_{ij} = \beta J_{ij}$ and $\beta = 1/T$.
The probability of a particular spin configuration \( \{S_i\} \) is given by
\begin{equation}
P(\{S_i\}) = \frac{1}{Z} \exp\left( \sum_{\langle ij \rangle} \tilde{J}_{ij} S_i S_j \right),
\end{equation}
where $Z$ is a normalization factor that ensures all probabilities sum to one (see Eq.~\eqref{eq:Z}). It contains the sum over all admissible configurations and encodes the total statistical weight of the system. In this framework, the most likely macroscopic state corresponds to the configuration that minimizes the energy landscape, justifying our use of the effective action as a variational principle in the subsequent analysis.

We decouple the interaction term by introducing an auxiliary field $\phi_i$ via a Hubbard-Stratonovich transformation
\begin{align}
Z &= \mathcal{N} \int \mathcal{D}\phi \sum_{\{S_i\}} \exp\left[ -\frac{1}{2} \sum_{ij} \phi_i \tilde{J}^{-1}_{ij} \phi_j + \sum_i S_i \phi_i \right] \nonumber  \\
&= \mathcal{N} \int \mathcal{D}\phi \exp\left[ -\frac{1}{2} \sum_{ij} \phi_i \tilde{J}^{-1}_{ij} \phi_j + \sum_i \log \sum_{\{S_i\}} e^{S_i \phi_i} \right].
\label{eq:ZHS}
\end{align}
In the spin-$1/2$ case, where $S_i = \pm 1$, the sum  over the configurations produces $\sum_{\{S_i\}} e^{S_i \phi_i} = 2 \cosh(\phi_i)$, leading to the effective action
\begin{equation}
S_{\mathrm{eff}}[\phi] = \frac{1}{2} \sum_{ij} \phi_i \tilde{J}^{-1}_{ij} \phi_j - \sum_i \ln(2 \cosh \phi_i).
\label{eq:action_complete}
\end{equation}

\paragraph*{\bf{Mean-field approximation}}
We now identify the homogeneous saddle point $\phi_i = \bar{\phi}$ by solving
\begin{equation}
\frac{\partial S_{\mathrm{eff}}}{\partial \phi_i} \bigg|_{\phi_i = \bar{\phi}} = 0
\quad \Rightarrow \quad
\bar{\phi} = \tilde{J}_0 \tanh(\bar{\phi}),
\label{eq:MF}
\end{equation}
where \( \tilde{J}_0 = \beta J_0 \), and \( J_0 = \sum_j J_{ij} \) is the zero-momentum component of $J_{\bf{q}} =  \sum_{r_{ij}} e^{- i \boldsymbol{q} (\boldsymbol{r}_i -  \boldsymbol{r}_j)} J_{i j}$. Notice that this identification only assumes that the coupling $J_{ij}$ is \textit{translation-invariant}, i.e., it depends only on the relative distance between sites: $J_{ij} = J(|\boldsymbol{r}_i - \boldsymbol{r}_j|)$.
Eq.(\ref{eq:MF}) reproduces the standard mean-field equation for the magnetization \( m = \bar{\phi} / \tilde{J}_0 \), allowing us to define the rescaled field $\phi \to \tilde{J}_0 \phi$.

Near the transition, \( \bar{\phi} \to 0 \), so the equation reduces to \( \bar{\phi}(1 - \tilde{J}_0) = 0 \), yielding a critical point at
\begin{equation}
T_0 = J_0.
\end{equation}
This relation holds independently of the specific lattice geometry; the latter enters only through $J_0$, which encodes the corresponding coordination and coupling structure.\\

\paragraph*{\bf{Fluctuation analysis}} To analyze the fluctuations around the mean-field value $\bar{\phi}$ we expand the action to quartic order. Near the critical point, where $\bar{\phi} = 0$, the effective action becomes
\begin{eqnarray}
S_{\mathrm{eff}}[\phi] &=& \frac{\tilde{J}_0^2}{2} \sum_{ij} \phi_i \tilde{J}^{-1}_{ij} \phi_j - \sum_i \bigg( \ln 2  +\nonumber \\
 &&+ \frac{1}{2} (\tilde{J}_0 \phi_i)^2  - \frac{1}{12} (\tilde{J}_0 \phi_i)^4 + \cdots \bigg),
\label{eq:action_expanded}
\end{eqnarray}
If the quartic coefficient becomes negative, the sixth-order term must be retained to ensure thermodynamic stability. A positive sixth-order coefficient guarantees that the free energy remains bounded from below and that a first-order transition can occur, as discussed in standard treatments of Landau theory~\cite{ChaikinLubensky, StrukovLevanyuk, Binder:1987aa, Wang2010}.

It is worth having a closer look at the quadratic term that in the momentum space reads 
\begin{eqnarray}
\label{eq:action2}
S^{(2)}_{\mathrm{eff}} &=& \frac{1}{2} \sum_{\boldsymbol{q}} \tilde{J}_0^2 \left( \tilde{J}_{\boldsymbol{q}}^{-1} - 1 \right) |\phi_{\boldsymbol{q}}|^2 \\
&& \frac{1}{2} \sum_{\boldsymbol{q}} \chi(\boldsymbol{q})^{-1} |\phi_{\boldsymbol{q}}|^2
\end{eqnarray}
where we defined the Gaussian susceptibility $\chi(\boldsymbol{q})$.\\

\paragraph*{\bf{Hypercubic lattice}} For a hypercubic lattice in $d$ dimensions with spacing $\ell_0$ and nearest-neighbor interactions $J_{ij} = J$, the Fourier transform is
\begin{equation}
J_{\boldsymbol{q}} = 2J \sum_{k=1}^d \cos(q_k \ell_0).
\label{J_q}
\end{equation}
Expanding at small $\bf{q}$ we find  
\begin{equation}
J_{\boldsymbol{q}}^{-1} \approx J_0^{-1} + \frac{\ell_0^2}{(2d)^2 J} \boldsymbol{q}^2.
\end{equation}
with \( J_0 = 2dJ \). Thus the quadratic part of the action, Eq(\ref{eq:action2}), reads
\begin{equation}
S^{(2)}_{\mathrm{eff}} = \frac{1}{2} \sum_{\boldsymbol{q}} \left[ a + \kappa \boldsymbol{q}^2 \right] |\phi_{\boldsymbol{q}}|^2,
\end{equation}
with
\begin{equation}
a = a_0 (T - T_0), \quad a_0 = \frac{2dJ}{T^2}, \quad T_0 = 2dJ, \quad \kappa = \frac{J \ell_0^2}{T}.
\label{eq:coefficientsIsing}
\end{equation}
Notice that the $\bf{q}=0$ part of the gaussian susceptibility $\chi(0) = a^{-1} \sim 1/(T-T_0)$ diverges at the transition, while the coefficient of the $\bf{q}^2$ term, $\kappa$ determines the rigidity of the order parameter.

The coarse-grained Landau action back into the real space reads
\begin{equation}
\label{eq:action_real}
\mathcal{S}_{\mathrm{eff}}[\phi] = \int d^d x \left[ \frac{a}{2} \phi^2(x) + \frac{\kappa}{2} (\nabla \phi)^2 + \frac{b}{4} \phi^4(x) \right],
\end{equation}
with the coefficient reported above and  
\begin{equation}
b = \frac{\tilde{J_o}^4}{3} = \frac{1}{3} \left(\frac{2dJ}{T}\right)^4  
\end{equation}.

\subsection{\bf{Generalization to spin-1 and Blume--Capel models}}

The same procedure described in the previos section applies to the spin-$1$ and Blume-Capel (BC) models. The core structure of the derivation remains unchanged, however, the set of allowed spin values now includes $S = 0$, which modifies the local sum in the partition function, Eq.(\ref{eq:ZHS}), and consequently alters the logarithmic term in the effective action, Eq.(\ref{eq:action_complete}).

For the spin-$1$ model, the local term becomes:
\begin{equation}
\label{eq:spin1-log}
\mathcal{S}_{\mathrm{loc}}^{(1)}(\phi) = -\log\left(2 \cosh \phi + 1\right),
\end{equation}
reflecting the additional entropy from the neutral state.

For the BC model, the configurations are further weighted by an energy $\Delta$, leading to:
\begin{equation}
\label{eq:BC-log}
\mathcal{S}_{\mathrm{loc}}^{\mathrm{BC}}(\phi) = -\log\left(2 e^{-\beta \Delta} \cosh \phi + 1\right).
\end{equation}

As shown in Eq.~\eqref{eq:action_expanded}, the expansion of these logarithmic terms around the saddle point determines the coefficients \( a \), \( \kappa \), and \( b \) in the effective theory. The presence of the neutral state increases the entropy and modifies the spin cumulants, leading to a reduced critical temperature \( T_0 \) and smaller quartic coefficient \( b \). In the BC case, these effects are further tunable via \( \Delta \). \\

\subsection{\bf{Effects of $\Delta$ in the Blume--Capel model}}
As we will see in the BC model the quartic coefficient of the effective action can become negative for moderate positive $\Delta$, signaling that the $\phi^4$ truncation is no longer sufficient to ensure stability of the functional. Since the Hamiltonian is $\mathbb{Z}_2$-symmetric, odd powers are absent, and the leading stabilizing contribution is the sixth-order term. For this reason we include already this term in the discussion.

Expanding Eq.(\ref{eq:BC-log}) around $\phi=0$ one finds (up to an irrelevant constant)
\begin{eqnarray}
\mathcal{S}_{\mathrm{loc}}^{\mathrm{BC}}(\phi) &=&
-\frac{1}{e^{\beta\Delta}+2}\,\phi^2
-\frac{e^{\beta\Delta}-4}{12\,(e^{\beta\Delta}+2)^2}\,\phi^4 \nonumber \\
& &+\frac{-e^{2\beta\Delta}+26e^{\beta\Delta}-64}{360\,(e^{\beta\Delta}+2)^3}\,\phi^6
+\mathcal{O}(\phi^8).
\label{eq:BC_loc_phi6}
\end{eqnarray}
Following the same steps discussed for the spin-$1/2$ case, including the rescaling of the Hubbard--Stratonovich field $\phi\to \tilde J_0 \phi$, we can rewrite the Landau action in the real space analogous to Eq.(\ref{eq:action_real}) 
\begin{eqnarray}
\mathcal{S}_{\mathrm{eff}}[\phi] &=& \int d^d x \bigg[ \frac{a}{2} \phi^2(x) + \frac{\kappa}{2} (\nabla \phi)^2 \nonumber \\
& & + \frac{b}{4} \phi^4(x)  + \frac{u}{6} \phi^6(x) \bigg],
\label{eq:action6_real}
\end{eqnarray}
with 
\begin{align}
a(\tilde\Delta) &= \left(\tilde J_0^{-1}-\frac{2} {e^{\tilde\Delta}+2}\right) \tilde J_0^{2},
\label{eq:a_BC_methods}\\[0.25em]
b(\tilde\Delta) &= \,\frac{4-e^{\tilde\Delta}}{3\,(e^{\tilde\Delta}+2)^2} \tilde J_0^{4},
\label{eq:b_BC_methods}\\[0.25em]
u(\tilde\Delta) &= \frac{-e^{2\tilde\Delta}+26e^{\tilde\Delta}-64}{60\,(e^{\tilde\Delta}+2)^3}\tilde J_0^{6}.
\label{eq:u_BC_methods}
\end{align}
where $\tilde\Delta \equiv \Delta/T$ and again $\tilde J_0=J_0/T=(2dJ)/T$ for nearest-neighbor interactions on the hypercubic lattice. $\kappa$ is given by the small-$\bf q$ expansion of $J_{\bf q}$ as before, Eq.(\ref{eq:coefficientsIsing}).

\paragraph*{\bf{Critical behavior}} The quadratic coefficient controls the instability. 
As we pointed out already, the ${\bf q} = 0$ part of the gaussian susceptibility $\chi(0) = a^{-1}$ diverges at the transition. At $T_0$ then using Eq.(\ref{eq:a_BC_methods}) we can write 
\begin{equation}
\frac{T_0}{2dJ} - \frac{2}{e^{\Delta/T_0}+2} =0 
\label{eq:T0_BC}
\end{equation}
where we use explicilty that $\tilde{J_0}=2dJ/T$. This transcendental equation has no closed-form solution, but for small $\Delta/T_0 \ll 1$ we can expand and obtain an approximate expression for $T_0$
\begin{equation}
T_0(\Delta) = \frac{4dJ}{3}\left(1-\frac{\Delta}{4dJ}\right).
\label{eq:T0_smallDelta}
\end{equation}
Near the criticality we can write $a(T)\sim a_0(T-T_0)$. 
To define $a_0$, it is convenient to separate in Eq.~\eqref{eq:a_BC_methods} the critical bracket from the scaling prefactor:
\begin{equation}
a(T)={\cal A}(T)\,\tilde J_0^2,
\qquad 
{\cal A}(T)\equiv \tilde J_0^{-1}-\frac{2}{e^{\Delta/T}+2}.
\end{equation}
By construction, ${\cal A}(T)$ is the factor that controls the instability: it vanishes as $T\to T_0$ and changes sign across the transition, whereas the rescaling factor $\tilde J_0^2$ is a smooth, non-vanishing prefactor that can be evaluated at $T_0$. 
With this convention we identify
\begin{equation}
a_0 \equiv {\cal A}'(T_0)\,\tilde J_0^2
= \left( \frac{1}{2dJ} -  \frac{2 \Delta \, e^{\Delta/T_0}}{T_0^2 (e^{\Delta/T_0}+2)^2} \right)\tilde J_0^2.
\end{equation}
For small $\Delta$, using $T = T_c^{(0)}=4dJ/3$ we find the leading correction to $a_0$
\begin{eqnarray}
a_0 &\simeq& \frac{\tilde{J_0}^2}{2dJ} \left(1 - \frac{\Delta}{4dJ}\right) \nonumber \\
&=& \frac{2dJ}{T^2} \left(1 - \frac{\Delta}{4dJ}\right) 
\end{eqnarray}
where we use the explicit definition of $\tilde{J_0}$. Notice that $a_0 = 2dJ/T^2$ at $\Delta=0$, continuously connecting to the spin-1 result.

The fourth order coefficient, defined in Eq.(\ref{eq:b_BC_methods}), explicitly reads
\begin{eqnarray}
b(\Delta) &=& \,\frac{4-e^{\Delta/T}}{3\,(e^{\Delta/T}+2)^2} \left(\frac{2dJ}{T}\right)^{4} \nonumber \\ 
\nonumber \\
&\sim&  \frac{1}{9}\bigg(\frac{2dJ}{T}\bigg)^4 \bigg( 1 - \frac{\Delta}{T}\bigg)
\end{eqnarray}
thus the lower-order correction in $\Delta$ is again linear. In the main text we use a compact notation to discuss the sign change of $b(\Delta)$ by defining a function ${\cal{B}}(\Delta/T)= (2dJ)^4/ (3 T^4\,(e^{\Delta/T}+2)^2)$ that is positive defined for any $\Delta$ and thus does not affect the sign of $b(\Delta)$.

\begin{figure}[tbh]
\centering
\includegraphics[width=0.95\linewidth]{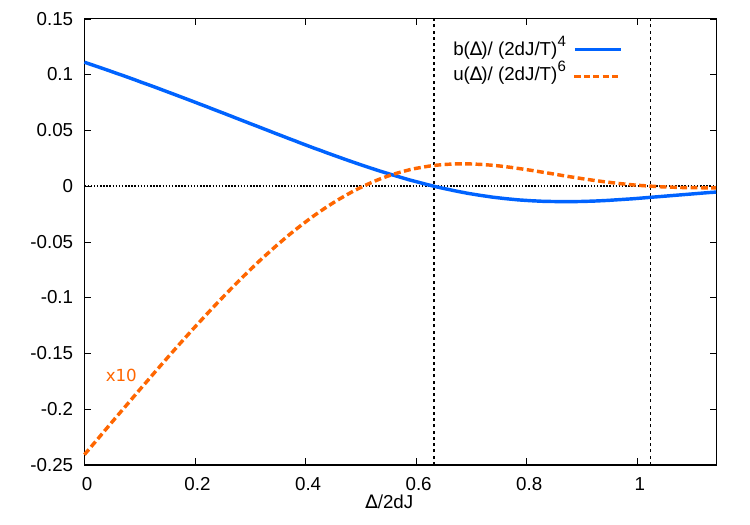}
\caption{Quartic and sixth-order Landau coefficients $b(\Delta)$ and $u(\Delta)$ of the Blume--Capel effective action, shown in the dimensionless form $b/(2dJ/T)^4$ and $u/(2dJ/T)^6$. For $b>0$ the $\phi^4$ truncation is stable. The sign change of $b$ signals the onset of a $\phi^6$-stabilized regime ($b<0$, $u>0$), associated with a first-order transition beyond the tricritical point. Beyond the second threshold where $u$ changes sign, the $\phi^6$ truncation no longer stabilizes the functional and higher-order terms (e.g. $\phi^8$) are required. The sixth-order curve is multiplied by $10$ for clarity.}
\label{fig:bu}
\end{figure}

\paragraph*{\bf{First order transition and tricritical point}}
We analyze now the nature of the phase transition. Since the denominators in Eqs.~\eqref{eq:b_BC_methods}--\eqref{eq:u_BC_methods} are strictly positive, the sign changes are governed by the numerators. The quartic coefficient changes sign when
\begin{equation}
b(\tilde\Delta)=0 \quad\Rightarrow\quad \tilde\Delta^\star=\ln 4,
\label{eq:Delta_star_methods}
\end{equation}
so that $b<0$ for $\tilde\Delta>\ln 4$. In this case stability of the truncated functional, Eq.(\ref{eq:action6_real}), requires $u>0$.
From Eq.~\eqref{eq:u_BC_methods} one finds that
\begin{equation}
u(\tilde\Delta)=0 \quad\Rightarrow\quad \tilde\Delta_{u,\pm}=\ln\!\left(13\pm\sqrt{105}\right),
\label{eq:Delta_u_methods}
\end{equation}
so that $u(\tilde\Delta)>0$ for $\tilde\Delta_{u,-}<\tilde\Delta<\tilde\Delta_{u,+}$. Since $\tilde\Delta^\star=\ln 4$ lies within this interval, the $\phi^6$ term stabilizes the Landau functional in the regime where $b<0$, up to $\tilde\Delta<\tilde\Delta_{u,+}$. For $\tilde\Delta>\tilde\Delta_{u,+}$ one has $u<0$ and higher-order terms (e.g. $\phi^8$) are required for stability.

The tricritical point, $T_{\rm TCP}$, corresponds to the simultaneous conditions $a=0$ and $b=0$ (with $u>0$). Imposing $a(\tilde\Delta^\star=\ln 4)=0$, from Eq.~\eqref{eq:a_BC_methods} we find
\begin{equation}
\tilde J_0^{-1}=\frac{2}{e^{\tilde\Delta^\star}+2} \quad \Ra \quad  \tilde J_0^{-1} = \frac{1}{3},
\end{equation}
Using that $\tilde J_0=(2dJ/T)$ for the hypercubic lattice, we obtain
\begin{equation}
T_{\rm TCP}=\frac{2 d J}{3}.
\label{eq:beta_TCP_methods}
\end{equation}
$T_{\rm TCP}$ is the boundary between a second and first order transition. Above $T_{\rm TCP}$ the onset of order is continuous (a graded change), whereas below it the system can switch discontinuously between competing minima (a jump-like transition), consistent with the need for a $\phi^6$ term to stabilize the Landau landscape when $b<0$.\\

Although the stability analysis is most naturally expressed in terms of the dimensionless control parameter $\tilde\Delta=\Delta/T$, it can be useful to re-express these results in terms of the microscopic anisotropy $\Delta$, e.g. to produce phase-diagram plots. Assuming $T \simeq T_0(\Delta)$, and using the definition of Eq.(\ref{eq:T0_smallDelta}), one can rewrite $\tilde{\Delta}$ in terms of $\Delta$
\begin{equation}
\tilde\Delta(\Delta)=\frac{\Delta}{T_0(\Delta)}=\frac{3\Delta}{4dJ-\Delta}.
\label{eq:tildeDelta_map_methods}
\end{equation}
Inverting we find
\begin{equation}
\Delta(\tilde\Delta)=\frac{2\tilde\Delta}{3+\tilde\Delta}\,(2dJ).
\label{eq:Delta_of_tildeDelta_methods}
\end{equation}
Evaluating this expression at the roots $\tilde\Delta^\star=\ln 4$ and $\tilde\Delta_{u,+}$ yields $\Delta^\star\simeq 0.632\, (2dJ)$ and $\Delta_{u,+}\simeq 1.024\, (2dJ)$. The latter sets the upper boundary of the $\phi^6$-stabilized regime relevant once $b<0$.
The results are summarized in Fig.~\ref{fig:bu}, where we plot the normalized coefficients $b(\Delta)$ and $u(\Delta)$ as functions of $\Delta$. The vertical lines indicate the thresholds at which $b$ and $u$ change sign. For $\Delta<\Delta^\star$, the $\phi^4$ truncation is stable ($b>0$) and the transition is second order, with a continuous onset of the order parameter. For $\Delta^\star<\Delta<\Delta_{u,+}$, one finds $b<0$ and $u>0$: the effective potential is stabilized by the $\phi^6$ term and the transition becomes first order. For $\Delta>\Delta_{u,+}$, $u<0$ and stability would require retaining higher-order even terms in the Landau expansion, which is beyond the scope of the present analysis.

\subsection{Generalization to more complex interaction structure}
\label{App:lattice}
Our analysis is robust with respect to moderate changes in the interaction structure. Within the present Landau-level treatment, modifying the $J_{ij}$ structure does not alter the form of the Hubbard--Stratonovich decoupling or the local $\phi$-expansions, but only renormalizes the coefficients through the interaction kernel (e.g. via $J_0$ and the small-$\mathbf q$ expansion).

To illustrate this point with a concrete but simple example, we includes an additional weak ferromagnetic next-nearest-neighbor coupling $J_d>0$ along the diagonal bonds, i.e. connecting ${\bf r}$ to ${\bf r}\pm \hat e_i \pm \hat e_j$ with $i\neq j$. The Hubbard--Stratonovich decoupling, the local single-site trace, and the resulting $\phi$-expansions are unchanged;
only the interaction kernel $J_{ij}$ (and hence $J_{\bf q}$, $J_0$, and the gradient stiffness) are modified.

On a $d$-dimensional hypercubic lattice with lattice spacing $\ell_0$ one finds
\begin{equation}
J_{\bf q}
= 2J \sum_{k=1}^d \cos(q_k \ell_0)
+ 4J_d \sum_{1\le l<m\le d} \cos(q_l \ell_0)\cos(q_m \ell_0),
\end{equation}
so that the uniform component becomes
\begin{equation}
J_0 \equiv J_{\bf q=0} = 2dJ + 2d(d-1)J_d .
\end{equation}
Expanding for small ${\bf q}$ gives
\begin{eqnarray}
J_{\bf q} &\simeq & J_0 - \ell_0^2 \Big[J + 2(d-1)J_d\Big]\,q^2 + \mathcal{O}(q^4),
\end{eqnarray}
i.e. the gradient term is stiffened by $J_d$ through the replacement $J \to J+2(d-1)J_d$ in $\kappa$ (cf. Eq.~\eqref{eq:coefficientsIsing}).

Since the local action is unaffected, the dependence on the microscopic interaction enters only through
$\tilde J_0 \equiv J_0/T$ and the small-$q$ expansion of $J_{\bf q}$.
In particular, the instability scale is set by the same conditions as before with the replacement $J_0 \to J_0(J,J_d)$
(e.g. for spin-$1/2$, $T_0=J_0$). Likewise, the Landau coefficients inherit the same overall scalings $a\propto \tilde J_0^{2}$, $b\propto \tilde J_0^{4}$, $u\propto \tilde J_0^{6}$, and the tricritical temperature remains $T_{\rm TCP}=J_0/3$ within this Landau-level treatment. Thus adding next-nearest-neighbor  couplings produces quantitative shifts of the Landau parameters without changing the qualitative $\phi^4/\phi^6$ stability structure discussed in the main text.

\begingroup
\sloppy
\bibliography{Landau.bib}
\endgroup

\end{document}